\documentclass[12pt,preprintnumbers,  letterpaper]{article}
\usepackage{color}

\usepackage{amsmath, amssymb, graphics, epsfig, graphicx, epstopdf}

\usepackage{amsfonts}
\usepackage{amssymb}
\usepackage{graphicx}
\usepackage{amstext}

\newcommand{\nc}{\newcommand}
\nc{\ba}{\begin{eqnarray}}
\nc{\ea}{\end{eqnarray}}
\nc{\nn}{\nonumber}
\newcommand{\bea}{\begin{eqnarray}}
\newcommand{\eea}{\end{eqnarray}}
\newcommand{\K}{|\bfk+\bfq|}
\nc{\be}{\begin{eqnarray}}
\nc{\ee}{\end{eqnarray}}

\nc{\bfk}{{\bf k }}
\nc{\bfQ}{{\bf Q}}
\nc{\bfx}{{\bf x }}
\nc{\pfp}{{\bf{p}}}
\nc{\bfp}{{\bf{p}}}
\nc{\bfq}{{\bf{q}}}
\nc{\tbf}{\textbf}
\nc{\de}{{\rm d}}
\nc{\calP}{  { \cal P} }
\nc{\calR}{  { \cal R} }
\nc{\im}{ \mathrm{Im} }
\nc{\sg}{ \mathrm{sgn} }
\nc{\smm}{{{}_-}}
\nc{\spp}{{{}_+}}
\begin{document}

\begin{flushright} {\footnotesize YITP-17-75}  \end{flushright}

\begin{center}

\def\thefootnote{\fnsymbol{footnote}}
{\Large  Bubble nucleation and inflationary perturbations 
}
\\[0.5cm]

{  Hassan Firouzjahi\footnote{firouz@ipm.ir }$^{1,2}$},
{ Sadra Jazayeri\footnote{sadraj@ipm.ir }$^{1,2}$},
{  Asieh Karami\footnote{karami@ipm.ir}$^{1}$},
{  Tahereh Rostami \footnote{t.rostami@ipm.ir }$^{1}$}

{\small \textit{$^{1}$ School of Astronomy, Institute for Research in Fundamental Sciences (IPM) \\ P.~O.~Box 19395-5531, Tehran, Iran }}\\
{\small \textit{$^{2}$ Yukawa Institute for theoretical Physics,
 Kyoto University, Kyoto 606-8502, Japan }}\\

\end{center}

\vspace{.8cm}

\hrule \vspace{0.5cm}


\begin{abstract}
In this work we study the imprints of bubble nucleation  on primordial inflationary perturbations. We assume that the bubble is formed via the tunneling of a spectator field from the false vacuum of its potential to its true vacuum. We consider the configuration in which the observable CMB sphere is initially outside of the 
bubble.  As the bubble expands, more and more regions of the exterior false vacuum, including our CMB sphere,  fall into the interior of the bubble.   The modes which leave the horizon during inflation  at the time when the bubble wall collides with  the observable CMB sphere are affected the most. The bubble wall induces  non-trivial anisotropic and scale dependent corrections in the  two point function of the curvature perturbation.  The corrections in the curvature perturbation and the diagonal and off-diagonal elements of CMB power spectrum are estimated.  

\end{abstract}
\vspace{0.5cm} \hrule
\def\thefootnote{\arabic{footnote}}
\setcounter{footnote}{0}

\newpage
\section{Introduction}
\label{intro}

Bubble nucleation from quantum tunneling in field space has captured significant interests in theoretical cosmology  over the past decades. Starting with the seminal works of Coleman and collaborators \cite{Coleman:1977py, Callan:1977pt, Coleman:1980aw}, bubble nucleation  has played important roles in the development of inflationary models, either in old inflation \cite{Guth:1980zm, Sato:1980yn} or in new inflation \cite{Linde:1981mu, Albrecht:1982wi, Hawking:1981fz}. In addition,  
the idea of vacuum decay and  tunneling  play vital roles in studies  of landscape and multiverse scenarios.  For a review of vacuum decay and related references see \cite{Weinberg:2012pjx}.

The mechanism  of  Coleman-De Luccia (CL) tunneling as the initial stage of the cosmological inflation has been studied extensively  in the literature in the context of open inflation, see for example  \cite{open1, open2, open3, open4, Garriga:1996pg, Garriga:1997ht, Sugimura:2013cra}. In these scenarios, we have a de Sitter (dS) mother universe in a false vacuum and our inflationary patch has resulted from a tunneling out of this false vacuum towards a region of the inflation potential suitable for slow-roll inflation. The kinematical picture in these works was  that our observable Universe, measured by cosmic microwave background (CMB)  scales,  lies within the bubble which forms after the tunneling. Since the created Universe inside the bubble is an isotropic and homogeneous open Universe, the induced power spectrum on CMB (though different than the usual Bunch-Davies flat space power spectrum) is statistically isotopic and homogeneous.

In the present work  we study a different configuration in which the observable CMB sphere lies initially  outside of the bubble. While the bubble wall expands relativistically, the CMB scales come inside the bubble. To simplify the analysis here we consider the case in which it takes a relatively long time for the bubble to completely cover the CMB sphere. This is somewhat a  fine-tuned situation as  most of the observers in the Universe either do not have causal contacts with the bubble or they fall inside the bubble very quickly. In order to keep every thing as simple as possible, we consider a scenario containing two fields, the inflaton field and a spectator field, which do not interact with each other directly. 
The potential for the  spectator field has two minima with slightly different values of the potential. The dominant source of inflationary expansion is the inflaton field which is slowly rolling on its potential. While the inflaton field is slowly rolling, the spectator field tunnels from its false vacuum to its true vacuum, forming the CL bubble in an inflationary background. We are interested in the gravitational effects of the formed bubble on the inflaton fluctuations and the corrections in primordial curvature perturbations power spectrum. 

Observationally, the imprint of the bubble nucleation on CMB fluctuations  is an interesting question. 
Indeed, there are indications of anomalies in CMB maps looking for answers beyond the simple setup of a nearly Gaussian, nearly scale invariant and statistically isotropic primordial power spectrum. In particular, there are indications of power asymmetry in CMB maps as suggested by the Planck and the WMAP observations \cite{Ade:2013nlj, Ade:2015hxq, Eriksen:2003db, Eriksen:2006xr, Eriksen:2007pc, Gordon:2005ai}.  Intuitively speaking, this power asymmetry suggests a dipole-type asymmetry in primordial power spectrum with an amplitude of few percent. The statistical significance of these anisotropies are under debate. But, it is an interesting question if the inferred asymmetry has a primordial origin. This has captured interests in the past few years. In particular, the idea of defects such as domain walls, monopoles and cosmic strings have been employed  in  \cite{Jazayeri:2014nya, Firouzjahi:2016fxf, Jazayeri:2017szw}
to explain the primordial origins of the power asymmetry. Depending on the dimensionality of the defect, 
it may break translation invariance or rotation invariance,  generating  anisotropies in primordial power spectrum. In these works, the defects had no dynamics, i.e. the defects were stationary in a fixed inflationary background. They induce power asymmetries by modifying the background inflationary geometry. 
In the present work, we extend the motivation of \cite{Jazayeri:2014nya, Firouzjahi:2016fxf, Jazayeri:2017szw} by employing the mechanism of vacuum decay and bubble nucleation 
as a possible solution of primordial power asymmetry. In comparison to \cite{Jazayeri:2014nya, Firouzjahi:2016fxf, Jazayeri:2017szw}, the present work has the interesting feature that the bubble has dynamics. Indeed,  the effect of the bubble on inflationary power spectrum is mainly due to relative motion of the bubble wall with respect to our CMB sphere as described above.

\section{The setup }
\label{setup}

Here we preset our setup in some details. As mentioned above, we consider a system of two fields, the inflaton field $\phi$ and the spectator field $\psi$. They are not coupled to each other directly but they feel each other's presence  indirectly via gravity.  The potential for $\psi$ has two minima with small 
difference $\Delta V$ in their potential values. The dominant source of energy for the inflationary background  is the inflaton potential on which the inflaton field is slowly rolling. Originally,  the field $\psi$ is on its false vacuum, but  as in CL mechanism, it undergoes a tunneling from its false vacuum to its true vacuum,  causing the bubble formation. We assume that $\psi$ is very massive compared to Hubble expansion rate during inflation $H$, so it can not be excited during inflation and it will be locked in on its true vacuum after tunneling. A similar setup was also studied in \cite{Sugimura:2012kr, Sugimura:2011tk}
looking for corrections from bubble nucleation on primordial non-Gaussianities.

The formation of bubble is a quantum mechanical process which can also affect the dynamics of the inflaton field. The bubble wall divides the spacetime into the regions inside the wall (true vacuum) and outside the wall (false vacuum) in which the inflaton field has different rollings. In addition,  the formation of bubble modifies the background geometry which affects the evolution of the inflaton field. Correspondingly, 
the effects of bubble  on the inflationary perturbations at the time of bubble formation and during its  subsequent evolution are complicated. Here we consider a simplified picture in which the bubble wall has reached its attractor speed limit and we are far from the complicated quantum mechanical effects of bubble formation  on inflationary evolution.  More specifically, we consider the  case where the effects of non-Bunch Davies vacuum, induced by the physics of tunneling,  have been erased by the subsequent expansion, hence we are left with the classical effects of the bubble. In this picture, the main effects of the bubble is to modify the background geometry for the inflation field which can be treated classically. Intuitively speaking, we have two distinct dS backgrounds: the interior of the wall and its ambient exterior,  which are separated by the bubble wall. As the wall expands relativistically, more and more regions of the exterior background fall into the interior regions.  Our assumption is that the observable CMB sphere is outside the formed bubble. But as the bubble expands rapidly, it will hit the CMB sphere and can encompass the entire or parts of CMB sphere, depending on relative kinematical configuration.

\begin{figure}[t!]
\vspace{-1.5cm}
	\centering
	\includegraphics[scale=0.45]{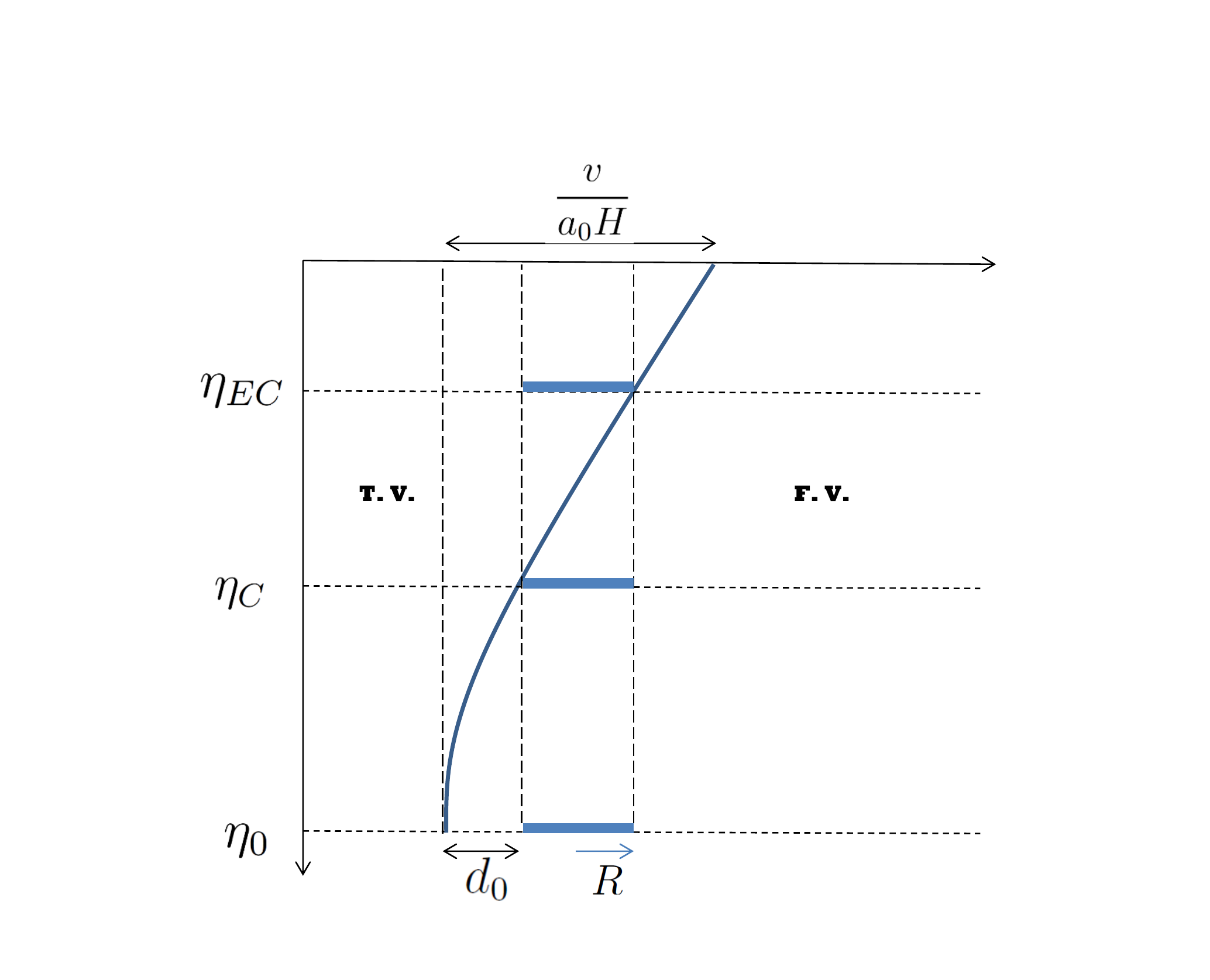}
	\includegraphics[scale=0.42]{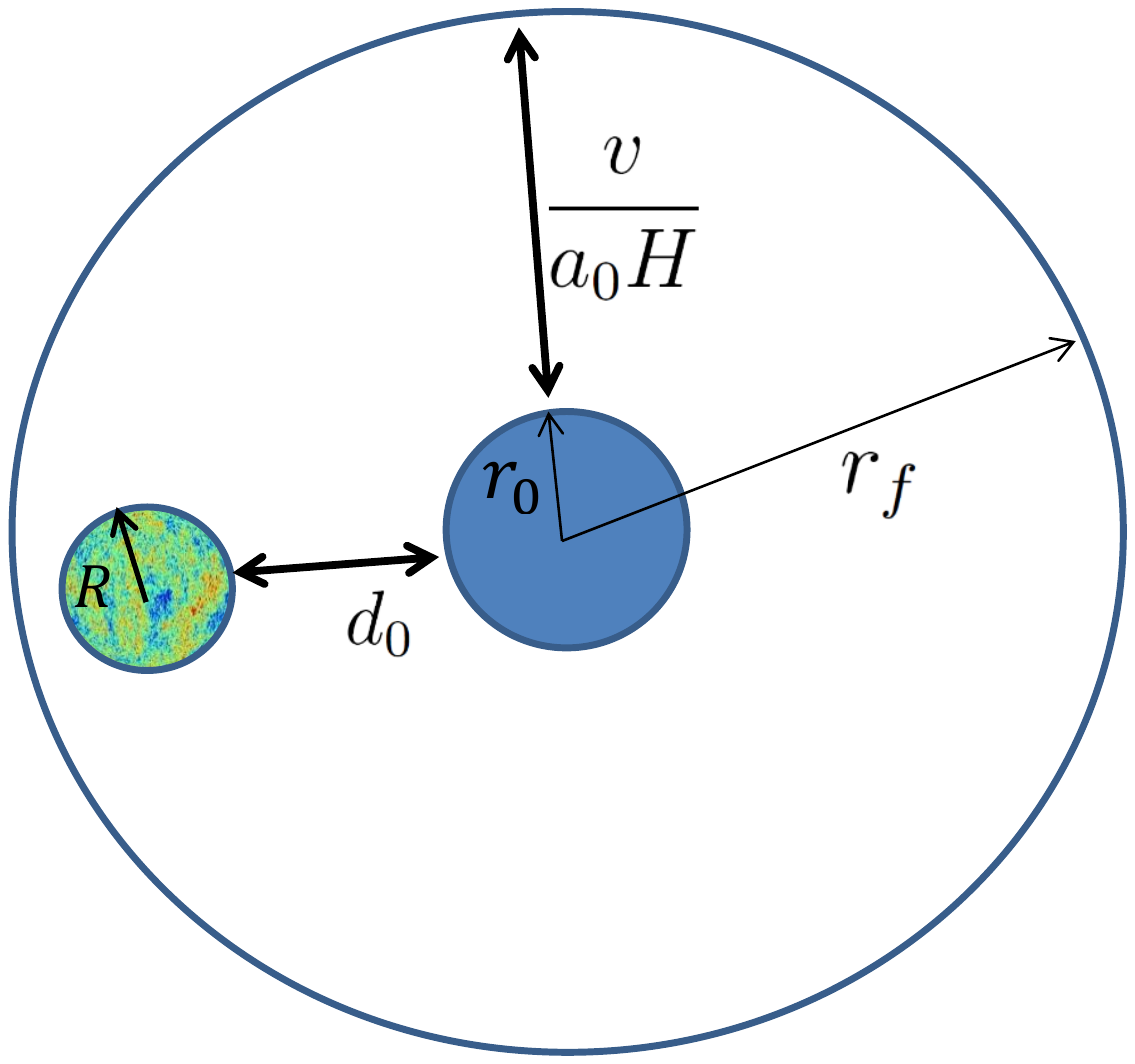}
	\caption{Left:  the diagram of bubble evolution in conformal time, in which the regions of true vacuum (T. V.) and false vacuum (F. V. ) are separated by the bubble wall, denoted by the curved solid line. 
	 Right: the evolution of the bubble with respect to the CMB sphere in spacetime.  $r_f$ is the final radius of the bubble in comoving scale.}
	\label{config}
\end{figure}

In Fig.  \ref{config} we have plotted the relative positions of the bubble at the time of formation  and the CMB sphere.  In this picture, $r_0$ is the comoving radius of bubble at the time of formation  $\eta_0$, $R$ is the comoving radius of the observed CMB sphere, $d_0$ is the distance between the center of CMB sphere and the center of bubble, $1/H$ is the radius of the Hubble patch and $r_f$ is the final comoving radius of the bubble.  We work in the coordinate system in which the origin is located at the center of the bubble and $\eta$ is the conformal time.

On the physical ground, the physical radius of the bubble can not exceed the Hubble radius which defines the causal patch in dS background,  so  $ a_0 r_0 < H^{-1}$, in which $a_0$ is the value of the scale factor at the time of bubble formation.  Looking at Fig.  \ref{config} the hierarchies $a_0 R <H^{-1}$ and $a_0 d_0 < v H^{-1}$ hold between  various radii in this configuration in which $v$ ($\simeq 1$)  is the asymptotic velocity of the bubble wall.   Correspondingly,  we define the following dimensionless kinematical parameters in our model
\ba
\label{parameters-def}
\beta\equiv a_0 r_0 H <1  \, , \quad 
\alpha \equiv\dfrac{1}{v} a_0 d_0 H<1\, , \quad  
\gamma\equiv \dfrac{2Ra_0H}{v}<1 \, .
\ea
Note that in the CL mechanism, the  physical size of the bubble at the time of formation is related to
the bubble tension $\sigma$ via $a_0 r_0=\frac{3\sigma}{\Delta V}$.

We have two important intermediate instants of time in our setup. $\eta_C$ is when the bubble first hits the CMB sphere on one side.   Assuming that the bubble has reached its asymptotic velocity, $v$, this time is given by
\be
\eta_C-\eta_0=\frac{d_{0}}{v} \, .
\ee
In terms of the number of e-folds $N$ (counted backward from the time of end of inflation when $N=0$, so $N \ge 0$ during inflation), this gives
\be
\exp(N_C-N_0)=1-\alpha \, .
\ee
To trust our assumption of neglecting the quantum mechanical effects of bubble formation on the inflaton dynamics, and the assumption that the bubble has enough time to reach its asymptotic velocity $v$ while approaching toward CMB sphere,  we require that $ N_0-N_C\sim  \mathrm{few} $. For example taking 
$ N_0-N_C=4$, requires a fine-tuning  at the order of one percent on $\alpha$, i.e. $\alpha \sim 0.99$. This corresponds to a configuration in which the boundaries of the bubble and CMB are almost a Hubble distance away(but still inside the Hubble patch).
As mentioned before, our assumption that  $\exp (N_0-N_C)\gg 1$ ensures that  we can neglect  the deviation from the Bunch-Davies vacuum induced by bubble nucleation.

The other important time, $\eta_{EC}$,  is the moment when the  bubble encompasses the entire CMB sphere, i.e. the time of end of bubble ``collision'' with CMB sphere.  The physical  expectation is that the  modes which exit the Hubble horizon after this moment do not feel the effect of bubble formation.  In terms of our  parameters, we have 
\be
\exp(N_{EC}-N_{C})=1-\dfrac{\gamma}{1-\alpha} \, .
\ee

Denoting the number of e-folds when the largest observable modes in CMB have crossed the Hubble horizon by $N_* \sim 60$, we have 
\ba
N_0 &=& N_*-\ln \gamma\, ,\\ \nn
N_C &=& N_*+\ln \dfrac{1-\alpha}{\gamma} \, ,\\ \nn
N_{EC} &=& N_*+\ln \dfrac{1-\alpha-\gamma}{\gamma} \, .
\ea

Depending on model parameters, there are two distinct possibilities: 

\begin{itemize}
\item $1-\alpha>\gamma$ so $N_C > N_*$.   If we require that  $N_{EC}$ to lie within the CMB observable window we need to  have $N_{EC}<N_*$ or $\dfrac{1-\alpha}{2}<\gamma$. The number of e-folds during which the CMB modes have been affected by  the formation of bubble is given by $N_*-N_{EC}=\ln \Big(\dfrac{1-\alpha}{\gamma}-1\Big )$ which is at the order of  1, unless we apply further fine-tuning on $\gamma$. 

\item $1-\alpha<\gamma$, so $N_C > N_*$.  The above formula for $N_{EC}$ does not apply; instead we have $N_{EC}=0$.
In this case  the CMB sphere never completely enter inside the bubble.  All modes which leave the horizon after  $N_C $ e-folds are affected by the expansion  of bubble wall. 
\end{itemize}

\section{Bubble wall expansion}
\label{wall-expansion}

As described above, the bubble wall expands relativistically after formation, dividing the spacetime into two 
FRW backgrounds. The interior is filled with the true vacuum while the exterior regions are still in false vacuum.  In the limit that we neglect the subleading slow-roll corrections, the interior and the exterior regions can be approximated by dS spacetimes with constant Hubble expansion rates $H_-$ and $H_+$.  We consider the situation where the difference in the values of the potential at the true and the false vacuum is small, $\Delta V \ll V_{inf}$,  in which $V_{inf}$ is the inflationary potential yielding the background (exterior region) expansion $V_{inf}= 3 M_P^2 H_+^2 $. 

Our goal in this section is to determine the dynamics of the expansion of the bubble wall into the false vacuum. This is required for our purpose to determine what an observer outside the  bubble (false vacuum) expanding with a time-like trajectory sees when swept  by the bubble, entering the true vacuum inside the bubble.   To answer this, first we need to know how the  bubble affects the background equations. Then, we will investigate the correction terms due to bubble as a perturbation to the background equations of motion.
For related works dealing with the expansion of bubble in cosmological backgrounds and its cosmological implications  see \cite{Blau:1986cw, Berezin:1987bc, Berezin:1982ur,Aguirre:2005xs, Aguirre:2005nt, Aguirre:2007an, Lake:1984pn, Garriga:2015fdk, Deng:2016vzb, Garriga:2006hw, Ansoldi:2014hta, Chen:2015ibc}.

Suppose the bubble wall, denoted by the time-like hypersurface $\Sigma$, divides the whole spacetime V
into two regions $V^{+}$, the exterior of bubble,  and $V^{-}$, the interior of bubble.  Let $n_{\mu}$ denotes the unit normal to $\Sigma$. To have a consistent solution of Einstein's field equations, we require that the induced metric on the three-dimensional hypersurface $\Sigma$ to be continuous while the extrinsic curvature $K_{\mu \nu} = n_{\mu;  \nu}$ on $\Sigma$ satisfies the Israel's junction condition \cite{Israel:1966rt}
\be
[K_{ij}]=-8\pi G\left(S_{ij}-\frac{1}{2}h_{ij} S\right),
\ee
where $G= 1/8 \pi M_P^2$ is the Newton constant,  $h_{ij}$ denotes the induced $3$-metric on $\Sigma$,  $S_{ij}$ representing the surface energy density with $S$ being its trace and the symbol $[\, ]$ denoting the discontinuity across $\Sigma$. For our bubble wall with surface energy density $\sigma$, we have
\ba
S_{ij} = - \sigma h_{ij} \, ,\quad  \quad S= - 3\sigma \, .
\ea

For the line elements on the exterior region $V^+$ and the interior region $V^-$, we take them to be dS spacetimes with the metrics
\be
ds^2 = -dt^2_\spp +a_\spp ^2(t_\spp ) \left( dr_\spp ^2+ r_\spp ^2 d\Omega^2 \right) ,
\ee
and
\be
ds^2 = -dt^2_\smm +a_\smm ^2(t_\smm )\big ( dr_\smm ^2+r_\smm ^2d\Omega^2  \big )\, ,
\ee
in which $d\Omega^2 = d \theta^2 + \sin^2 \theta \, d \phi^2 $ represents the angular part of the metric which is the same in both spacetimes.  Note that we have allowed the coordinates $(t, r)$ to be different on  $V^{\pm}$, as their relations will be fixed after imposing the junction conditions. 

The induced metric on the bubble wall $\Sigma$ has the form 
\bea
ds^2 = -d\tau^2+R^2(\tau)d\Omega^2 \, ,
\eea
where $R(\tau)$ represents the radius of the bubble as measured by the time $\tau$ accessible to the observer confined on $\Sigma$. 

The unit normal $n_\mu$ is given by 
\ba
n_{\mu}=\left(\mp a(t_\pm) \, r_\pm',\pm a(t_\pm)\,  t_\pm' ,0,0\right) \, ,  
\ea
where a prime indicates the derivative with respect to $\tau$. 

The conditions of the continuity of the metric on $\Sigma$ require that 
\bea
\label{con1}
R(\tau)= R_+(\tau) = R_- (\tau) \, ,
 \eea
 where $R_\pm \equiv a(t_\pm) r_\pm$,  and 
 \be\label{con}
 d\tau^2 =dt_\spp ^2- a(t_+)^2 dr_\spp ^2
 = dt_\smm ^2-a_\smm ^2(t_\smm )dr_\smm ^2 \, .
 \ee
 The latter equation yields
 \ba
 \label{Cond}
{ t_{\pm}'}^2 -  a(t_\pm)^2 { r_{\pm}'}^2 =1 \, .
 \ea

The relevant component of extrinsic curvature is given by
\ba
K_{\theta \theta}^\pm = R \left( a(t_\pm)^{-1} \frac{\partial R}{\partial r_\pm} t_\pm' + a(t_\pm) R' r_\pm'  \right)\, . 
\ea 
 Now, using the condition (\ref{Cond}) we obtain 
\ba
\label{K2}
(K_{\theta\theta}^{\pm })^{2}=R^2\left( {R'}^2+(1-{2m_{\pm}\over R})\right) \, ,
\ea
where we have defined the Misner-Sharp mass  \cite{Misner:1964je, McVittie:1933zz}  as
\ba
\label{misner}
m_{\pm} &=& \frac{R_\pm}{2G}\left(1-g_\pm^{\mu\nu}\partial_{\mu}R_\pm\partial_{\nu}R_\pm \right) \nonumber \\
&=& \frac{R^3H_\pm^2}{2G} \, . 
\ea

Finally from equations (\ref{K2}) and (\ref{misner}), the equation of motion of the shell is obtained to be 
\ba
\label{EOM}
{R'}^2+1=\left(\frac{m_\spp -m_\smm }{4 \pi \sigma R^2}\right)^2+\left(\frac{m_\spp +m_\smm }{R}\right)+4 \pi^2 \sigma^2 R^2  \, .
\ea
The above equation can be written as
 \ba
 \label{shell-eq}
 A^2R^2-R'^2=1 \,  ,
\ea
 where  we have defined 
\ba
\label{A-def}
A^2 \equiv\frac{\sigma^2}{16M_P^4}+\frac{H_+^2+H_{-}^2}{2}+\frac{M_P^4}{\sigma^2} {(H_+^2-H_-^2)^2} \,  .
\ea 

In our picture, the difference between the two Hubble expansion rates $H_{\pm}$ is small so we define
$H_{+} = H_{-}(1+ \epsilon)$ in which $\epsilon$ is a small dimensionless parameter. Correspondingly, the difference between two minima $\Delta V$ is given by 
\ba
\Delta V= 3 M_{P}^{2} (H_{+}^{2 } -  H_{-}^{2 } )= 6 M_{P^{2}} H \epsilon H = 2 \epsilon V \, .
\ea
On the other hand, the initial physics size of the bubble is related to $\Delta V$ via  $a_{0} r_{0} = 3 \sigma/\Delta V$ \cite{Coleman:1977py}. This can be used to express $\sigma$ in terms of the dimensionless parameter $\beta$ defined in Eq. (\ref{parameters-def}) as   
\ba
\label{beta-sigma}
\sigma=2\epsilon M_p^2H_-^2 \beta \, .
\ea  
Plugging this relation in the definition of parameter $A$ defined in Eq. (\ref{A-def}) yields 
 \ba  
 \label{A-eq}
 A^2=H_-^2 \Big(\frac{1}{\beta^2}+(1+\epsilon)+\epsilon^2\beta^2\Big) \, .
\ea
In our analysis below, we express the physical results to linear order in $\epsilon \ll 1$. 

Before proceeding to next section, there is one comment in order.
In writing the metric of the exterior region $V^+$ we have assumed that the bubble has no effective mass.
Otherwise, the metric of the exterior region should be in the form of dS-Schwarzschild solution. On the physical ground, we do not expect this to be the case. The simple reason is that the regions far from the bubble shell have not felt the formation of the bubble. Therefore,  based on causality, the metric in the region $V^+$ can not be affected by the shell and they should retain their original dS metric with the Hubble expansion rate $H_+$.  Technically speaking,  based on energy conservation, the negative energy inside the bubble is canceled by the positive tension of the bubble's wall. This cancellation is exact in Minkowski spacetime  and is expected to hold to leading order in $M_P^{-2}$ \cite{Coleman:1980aw}.

\section{The relations between two coordinates}
\label{relation}

In this section, we find the relation between the coordinates $(t_-, r_-)$ and   $(t_+, r_+)$ for the interior and the exterior regions of the shell. 

The solution of shell expansion from Eq. (\ref{shell-eq}) is given by 
  \be
  R(\tau)=\frac{1}{A}\cosh (A\tau) \, ,
  \ee
in which we have absorbed an unimportant phase by shifting the origin of $\tau$. 
  
Now, imposing the continuity of the induced metric on hypersurface $\Sigma$ as given in 
Eqs. (\ref{con1})  and (\ref{Cond}), we require 
\ba
\label{exact}
\exp(H_\pm t_\pm)r_\pm(t_\pm) &=&\frac{1}{A}\cosh (A\tau)\, ,  \nn \\
t_\pm'^2-\exp (2H_\pm t_\pm)r_\pm'^2 &=&1 \, .
\ea
From these equations  we obtain
\ba
\frac{H_\pm}{A}t_\pm'+\frac{r_\pm'}{A\,  r_\pm}=\tanh(A\tau) \nn , \\ 
t_\pm'^2-\frac{r_\pm'^2}{A^2 r_\pm^2}\cosh^2(A\tau)=1 \, .
\ea
These equations  further  yield 
\be
t_\pm'^2 \Big(1-\frac{H_\pm^2}{A^2}\cosh^2(A\tau)\Big)+\frac{H_\pm}{A}\sinh (2A\tau)t'_\pm=\cosh^2(A\tau) \, .
\ee  
The branch of solution with the requirement $t'>0$  is
\be
\label{tp}
t_\pm'=\frac{\frac{H_\pm}{A}\sinh (A\tau)- \sqrt{1-\frac{H_\pm^2}{A^2}}}{\frac{H_\pm^2}{A^2}\cosh^2(A\tau)-1}\cosh (A\tau) \, .
\ee
  
As mentioned in the previous section, we are interested in the asymptotic limit when sufficient time has elapsed since the bubble is formed and the bubble wall  has reached its final velocity.  Therefore, we solve the above equation perturbatively in powers of  $\exp(-A\tau)$. Defining  $y \equiv \exp(A\tau)$,  the asymptotic solution of Eq. (\ref{tp} ) to leading order for $y \gg 1$ is given by 
\be
\label{texact}
t(y)=\frac{1}{H}\ln(y)+2\sqrt{1-\frac{H^2}{A^2}}\frac{A}{H^2}y^{-1}+{\cal O}(y^{-2})+{ C} \, .
\ee
Putting $C=0$ on both sides of shell yields\footnote{By this choice, in principle the instant of the bubble formation is fixed and it does not necessarily corresponds to  $t_\pm=0$.}
\be
\label{tptm}
H_+t_+= H_-t_--\frac{2\epsilon}{\beta}(2+\beta^2)\exp(-H_-t_-)+{\cal O}(\exp(-2H_-t_-)) \, .
\ee
  
Now using 
  \be
  \nn
  r_\pm\exp (H_\pm t_\pm)=\frac{1}{A}\cosh (A\tau)=
  \frac{1}{2A}(y+y^{-1}) \, ,
  \ee
we can compute $r(y)$ for both interior  and exterior regions as follows 
\ba
\label{rexact}
r_\pm(y)=\frac{1}{2A}-\frac{\sqrt{1-\frac{H_\pm^2}{A^2}}}{H_\pm \, y}+{\cal O}(y^{-2}) \, .
\ea
From the above equation we see that asymptotically $r\to \frac{1}{2A}$.   

Using the above relation in addition to Eq. \eqref{texact}, one can read the asymptotic velocity of the bubble as follows 
\be
\label{vel}
v = a(t_\pm)\dfrac{dr_{\pm}}{dt_{\pm}}\to \frac{1}{\sqrt{1+\beta^2}}\pm\dfrac{1}{2}\dfrac{\beta^4 \epsilon}{(1+\beta^2)^{3/2}} \, .
\ee

Finally,  the relation between $r_+$ and $r_-$ (near $r_-\sim \frac{1}{2A}$ but not at the exact final point), to leading order is given by
\be
r_+=\Big (1- \frac{\epsilon}{1-\frac{H^2}{A^2}}\Big)r_-+\frac{\epsilon}{2A(1-\frac{H^2}{A^2})}+
{\cal O}\Big ((r_--\frac{1}{2A})^2\Big ) \, .
\ee
Using the formula for $A$ given in Eq. (\ref{A-eq}), we can express $r_-$ in terms of $r_+$ as follows 
\be
\label{rprm}
r_-=\Big(1+\epsilon (1+\beta^2)\Big)r_+-\frac{\epsilon}{2H}\beta \sqrt{1+\beta^2}+{\cal O}\Big ((r_+-\frac{1}{2A})^2\Big ) \, .
\ee

\section{Constraints from CMB low $\ell$ multipoles}
\label{low-l}

Our main interest is to find the corrections induced by bubble in curvature perturbations 
\ba
\label{R-eq}
\calR = -\frac{H}{ \dot \phi}  \delta \phi \, ,
\ea
 in which $\delta \phi$ is the inflaton perturbations. 

There are two types of corrections from bubble wall into curvature perturbations  power spectrum: the classical effect and the quantum effect. The classical effect originates from the fact that the formation of the bubble breaks the homogeneity of the spacetime, dividing it into the exterior and the interior regions. As a result, the classical evolution of the inflaton field will receive an $r$-dependent correction. This  
inhomogeneous correction in turn induces a correction in curvature perturbations. An extensive analysis of the shape and the amplitude of this classical correction is beyond the scope of this paper as one has to solve for the evolution of the inflaton field as a function of $r$ and $t$ for both the interior and the exterior regions of the bubble.   Here we provide  the order of magnitude estimation of the classical  correction in 
comparison to the leading  quantum corrections.  

Assuming an inhomogeneous $r$-dependence for the background inflaton field $\phi=\phi(r, t)$, the solution has to be smooth across the surface $\Sigma$, requiring the continuity of both $\phi$ and its normal derivative on $\Sigma$ as follows
\ba
\label{cont}
\phi(t_+(\tau),r_+(\tau))&=&\phi\left(t_-(\tau),r_-(\tau) \right) \, \\ 
\label{nor}
n^\mu\partial_\mu \phi|_+ &=& n^\mu \partial_\mu \phi |_- \, .
\ea
It is not easy to solve the Klein-Gordon equation, subject to above boundary conditions. Instead, as just mentioned, we provide an order of magnitude estimation of the dependence of the background solution on $r$.  Physically, we expect that for the exterior region $\phi_+=\phi_+(t_+)$, since because of the causality,  
the bubble formation does not affect the exterior  region. however,  we let   $\phi_-$ to depend to both 
$t_-$ and $r_-$. 

The normal to the bubble surface  in spherical coordinate reads
\ba
n_{\mu}=\dfrac{1}{\Big(\dfrac{1}{a^2}(\dfrac{dt}{d\tau})^2-(\dfrac{dr}{d\tau})^2\Big)^{\frac{1}{2}}} \left(-\dfrac{dr}{d\tau},\dfrac{dt}{d\tau},0,0 \right)=a \left(-\dfrac{dr}{d\tau},\dfrac{dt}{d\tau},0,0 \right) \, .
\ea
In the zeroth order in $\partial_r$, $\phi_-=\phi_-(t_-)$. Using \eqref{cont} and \eqref{tptm} we obtain
\be
\phi_-(t_-)=\phi_+\Big(\dfrac{H_-}{H_+}t_- \Big)+{\cal O}(\partial_r \phi_-) \, .
\ee
In addition,  Eq. \eqref{rprm} yields  
\ba
\dfrac{1}{a_-}\dfrac{dt_-}{d\tau}\partial_r \phi_- &\simeq& \Big(a_+\dfrac{dr_+}{d\tau}\partial_t \phi_+-a_-\dfrac{dr_-}{d\tau}\partial_t \phi_-\Big) \nn , \\ 
&\simeq& a_+r_+\Big(\dfrac{1}{r_+}\dfrac{dr_+}{d\tau}\partial_t \phi_+-\dfrac{1}{r_-}\dfrac{dr_-}{d\tau}\dfrac{H_-}{H_+}\partial_t \phi_+\Big) \, .
\ea
Using $\dfrac{1}{a}\dfrac{dt_-}{dr_-}\sim \dfrac{1}{\sqrt{1+\beta^2}}$, the above expression  further simplifies to 
\ba
\partial_r \phi_-\sim\sqrt{1+\beta^2}a\dot{\phi}\epsilon \Big(1-\dfrac{\beta}{2Hr}\sqrt{1+\beta^2}\Big) \, ,
\ea
in which a dot indicate a derivative with respect to cosmic time $t_+$. 

On the bubble  wall we have  $r\simeq \dfrac{1}{2A}\sim \dfrac{\beta}{2H\sqrt{1+\beta^2+\beta^2\epsilon}}$, so the above formula simplifies to 
\ba
\dfrac{1}{a}\partial_r \phi_-\sim -\dot{\phi}\epsilon \beta^2 \sqrt{1+\beta^2} \, .
\ea
The above expressions provides an estimation of the inhomogeneous corrections in the background evolution of $\phi$ calculated on the surface of the bubble. 

Though we have not calculated the profile of the inflaton field inside the bubble but from the above calculation we naively expect that the amplitudes of the induced quadrupole and octupole corrections in power spectrum of curvature perturbations Eq. (\ref{R-eq}) to be  of the same order as $\epsilon \beta^2$. In particular, if we expand the curvature perturbation  near the center of the CMB sphere, we have 
\be
\calR \simeq {\cal R}_0+d_i x^i+\dfrac{1}{2!}Q_{ij}x^ix^j+\dfrac{1}{3!}O_{ijk}x^ix^jx^k+...  .
\ee
The amplitudes of $Q_{ij}$ (quadrupole) and $Q_{ijk}$ (octupole) are constrained by the low $\ell$ multipoles of CMB. Roughly speaking we need $Q, O, \lesssim 10^{-5} $ \cite{Erickcek:2008sm}. (However note that $d_i$,  the  dipole part,  cancels from every observable \cite{Mirbabayi:2014hda,Erickcek:2008jp}). Therefore,  we conservatively conclude that the quadrupole and octupole constraints from 
CMB imply that $\epsilon \beta^{2 } \lesssim 10^{-5}$. In the realistic situation that $\epsilon$ is not 
unnaturally small, say taking $\epsilon \sim 10^{-2}$, we conclude that $\beta \ll 1$. Therefore, in the rest of our analysis  we work to leading order in 
$\beta$, discarding higher powers of  $\beta $.   A small $\beta$ corresponds to a small size of bubble at the time of formation. In this limit, according to \eqref{vel}, the bubble wall rapidly reaches the speed of light.

\section{Interaction Hamiltonian}
\label{Hamiltonian}
 
Our goal is  to calculate the corrections in curvature perturbations power spectrum induced by the formation of bubble. The logic is similar to  \cite{Jazayeri:2014nya, Firouzjahi:2016fxf, Jazayeri:2017szw} and \cite{Tseng:2009xw} where the imprints of various topological defects during inflation on curvature perturbations are studied, see also  \cite{Carroll:2008br, Prokopec:2010nm, Wang:2011pb, Cho:2009en, Cho:2014nka} for related works on defects during inflation.  In our picture, the bubble is essentially a defect with a surface energy density which is expanding in the exterior dS background. As in the case of topological defects, the bubble modifies the background geometry. This modification in geometry is felt by the inflaton field, inducing a correction in Hamiltonian which  leads to  modifications in  the curvature perturbation power spectrum. Our goal in this section is to calculate the interaction Hamiltonian to leading order in model parameters, namely $\epsilon$ and $\beta$.

As in  \cite{Tseng:2009xw, Jazayeri:2014nya, Firouzjahi:2016fxf, Jazayeri:2017szw},  we neglect the gravitational backreaction of the inflaton field on the background geometry. The gravitational backreaction of inflaton modifies the dS geometry at the order of slow-roll parameter $\epsilon_H = -\dot H/H^2$. Noting that $\epsilon_H \sim {\dot{\phi}^2}/{M_p^2 H^2}$, the combined contribution of slow-roll correction (i.e. gravitational backreaction of inflaton) and the bubble induces correction of order $\epsilon \sqrt \epsilon_H$ in curvature perturbation power spectrum.  In the slow-roll limit, $\epsilon_H \ll 1$, this can be neglected compared to direct contribution of bubble on dS geometry which will be at the order $\epsilon$.  

The metric outside of the bubble has its original form, i.e. the form before bubble formation. However, the metric of the interior is different than the exterior as was obtained in section \ref{relation}. To calculate the whole action, it is better if we express the metric of the interior region in terms of the exterior coordinates using the relations between the two coordinates obtained in section \ref{relation}. The metric of the interior of the bubble, expressed in terms of the exterior coordinate, is given by 
\ba
ds^2 &=&-\Big(\frac{dt_-}{dt_+} \Big)^{2}dt_+^2+\exp\Big(2H_-t_-(t_+) \Big)\Big [ \big(\frac{dr_-}{dr_+} \big)^{2}dr_+^2+ r_-(r_+)^2 d\Omega^2\Big ]\, ,\\ \nn
&=&(-1-2\epsilon)dt_+^{ 2}+\exp(2H_+t_+)\Big[\Big (1+2\epsilon (1+\beta^2)\Big )dr_+^{2}+r_-(r_+)^2 d\Omega^2\Big] \, .
\ea

Now the whole spacetime can be expressed in terms of the  exterior (original) coordinates $t_+, r_+$ as follows
\ba
ds^2=-dt^2+\exp (2H_+t)(dr^2+r^2d\Omega^2)+\delta g_{\mu\nu} \theta (t-t_0)\theta(R(t)-r) 
dx^\mu dx^\nu \, ,
\ea
where to simplify the notation we have removed the subscript $``+''$ so $t$ and $r$ actually mean $t_+$ and $r_+$ respectively in the following analysis.  

In this coordinate system, the correction in metric geometry induced by the bubble is given by 
\begin{align}
& \delta g_{00}=-2\epsilon \\ 
&\delta g_{rr}=2a^2\epsilon(1+\beta^2) \simeq 2a^2\epsilon
\\ 
&\delta g_{\theta\theta}=\sin^{-2}\theta \delta g_{\phi\phi}
\simeq  2a^2r^2\epsilon  \Big(1-\dfrac{\beta}{2Hr}  \Big)
\end{align}
In the third equation, we have used Eq. (\ref{rprm}) to express $r_-(r)$ in terms of $r_+$. Also we have discarded ${\cal O}(\beta^2)$ corrections as explained at the end of section \ref{low-l}.

In the new coordinate which is smooth over the bubble, the background inflaton field is homogeneous $\langle \phi\rangle=\phi(t)$. Neglecting the gravitational backreaction,  the interaction Hamiltonian for the inflaton perturbations is given by 
\ba
H_I=\theta(t-t_0)\int d^3x \Big (\sqrt{-g}\frac{1}{2}\delta g^{\mu\nu}\partial_\mu\delta \phi \partial_\nu \delta \phi+\delta \sqrt{-g} \frac{1}{2}g^{\mu\nu}\partial_\mu\delta \phi \partial_\nu \delta \phi\Big ) \, .
\ea
Now using
\ba
\delta \sqrt{-g}= \epsilon \sqrt{-g}\Big(6-\dfrac{\beta}{Hr} \Big) \, ,
\ea
the interaction Hamiltonian to leading order in $\epsilon$ and $\beta$ is obtained to be 
\ba
\label{HI-form}
H_I(t)=2\epsilon\theta(t-t_0)\int_0^{r_{W}(t)}a^3r^2 drd\Omega \Big [\dfrac{\delta \dot{\phi}^2}{2}(-1+\dfrac{\beta}{2Hr}) +\dfrac{(\nabla \delta \phi)^2}{ 2a^2}-\dfrac{\beta  (\partial_r \delta \phi)^2}{4 a^2Hr}
\Big ] \, ,
\ea
in which $t_{0}$ represents the time of the formation of bubble and 
$r_{W}(t)$ represents the time dependent comoving radius of bubble. In our approximation where the bubble has reached its final expanding stage, from Eq. (\ref{vel}) we have 
\ba
r_{W}(t) = r_{f} + v \eta \, .
\ea 
in which $\eta = -1/a H$ is the conformal time. 

In obtaining the interaction Hamiltonian Eq. (\ref{HI-form})  we have neglected the terms at the order   
$\epsilon \beta^2$  but we have kept terms  at the order $\epsilon \beta$. This is because inside the bubble $0<r<r_{W}(t)$, and the asymptotic value for  $r_{W}(t)$ is given by ${1}/{2A}\sim \beta/{2H}$, hence 
${\epsilon \beta  }/{Hr}>2\epsilon$.  Therefore,  this term is not negligible even near the surface of the bubble. 

\section{The effects of bubble on curvature perturbations}
\label{bubble-R}

Having obtained the interaction Hamiltonian, we are ready to calculate the corrections in the curvature perturbation two point function. Since $\calR$ is proportional to inflaton perturbations $\delta \phi$,  we need to calculate the corrections in inflaton two point function.  More specifically, 
with the curvature perturbation given in Eq. (\ref{R-eq}), the correction  in the curvature perturbation two point function induced by the bubble is 
\ba
\label{power}
\Delta \big \langle \calR_\bfk (t_e) \calR_\bfq(t_e)\big \rangle = \left( \frac{H^2}{\dot \phi} \right)^2 
\Delta \big \langle \delta \phi_\bfk (t_e) \delta \phi_\bfq(t_e)\big \rangle \, .
\ea


\subsection{In-In formalism}
\label{in-in}

Using the standard in-in formalism \cite{Maldacena:2002vr, Weinberg:2005vy}, the corrections in inflaton's 
two point correlation induced by bubble  to leading order in $\epsilon$ is  given by    
\ba
\label{Dyson}
\Delta \big \langle \delta \phi_\bfk (t_e) \delta \phi_\bfq(t_e)\big \rangle &=& i \int_{0}^{t_e} dt' \big \langle \left[ \, H_I(t'),  \, \delta \phi_\bfk \delta \phi_\bfq  \, \right ] \big \rangle  \nonumber\\
&=&- 2 \im \int_0^{t_e} d t' \big \langle   H_I(t') \delta \phi_{\bfk} (t_e) \delta \phi_{\bfq}(t_e) 
\big \rangle  \, ,
\ea
in which $t_e$ indicates the time of end of inflation.

To proceed further, we need to calculate $H_I$ in Fourier space. We present the Fourier transformation of the three types of terms in $H_I$ separately as follows: 
\ba
\int_{0}^{r_{W}(t)} r^2drd\Omega (\nabla \delta \phi)^2&=&\int \dfrac{d^3kd^3q}{(2\pi)^6}\dfrac{-4\pi\bfk.\bfq}{|\bfk+\bfq|^3}\delta \phi_\bfk\delta \phi_\bfq  \nn \\ 
&&\times \Big(\sin(|\bfk+\bfq| r_{W}(t))-r_{W}(t)|\bfk+\bfq|\cos (|\bfk+\bfq |r_{W}(t))\Big) , \\
\int_{0}^{r_{W}(t)} r^2drd\Omega (\delta \dot{\phi})^2 &=&\int \dfrac{d^3kd^3q}{(2\pi)^6}\dfrac{4\pi}{|\bfk+\bfq|^3}\delta \dot{\phi}_\bfk\delta \dot{\phi}_\bfq  \nn \\ 
&&\times \Big(\sin(|\bfk+\bfq|r_{W}(t))-r_{W}(t)|\bfk+\bfq|\cos (|\bfk+\bfq |r_{W}(t))\Big) , \\ 
\int_{0}^{r_{W}(t)} r drd\Omega (\delta \dot{\phi})^2&=&\int \dfrac{d^3kd^3q}{(2\pi)^6}\dfrac{4\pi}{|\bfk+\bfq|^2}\delta \dot{\phi}_\bfk\delta \dot{\phi}_\bfq \Big(1-\cos (|\bfk+\bfq|r_{W}(t))\Big) \, .
\ea

The Fourier transformation of the term $(\partial_r \delta \phi)^2$ is more involved, so before  integrating over $\bfk,\bfq$,  we use a new coordinate such that
\ba
\bfk+\bfq =|\bfk+\bfq|\hat{k} \, ,\quad 
\bfk = k\cos\psi \hat{k}+k\sin\psi \hat{j} \, , \quad 
\bfq = q\cos(\alpha-\psi)\hat{k}+q\sin (\psi-\alpha)\hat{j}\, ,
\ea
where $\alpha$ is the angle between $\bfk,\bfq$ while $\psi$ is the angle between $\bfk$ and $\bfk+\bfq$. It is easy to verify that 
\ba
\cos\psi=\dfrac{k+q\cos\alpha}{|\bfk+\bfq|}, \qquad 
\sin\psi=\dfrac{q\sin\alpha}{|\bfk+\bfq|}, \qquad
\cos (\alpha-\psi)=\dfrac{q+k\cos\alpha}{|\bfk+\bfq|} \, .
\ea

Now we are ready to compute the following integral 
\ba
\int_{0}^{r_{W}(t)}rdrd\Omega (\partial_r \delta \phi)^2&=&\int \dfrac{d^3kd^3q}{(2\pi)^6}\delta \phi_\bfk\delta \phi_\bfq \int_0^{r_{W}(t)}\dfrac{1}{r} drd\Omega (-\bfk.\bfx)(\bfq.\bfx)\exp(i(\bfk+\bfq).\bfx)\nn \\ 
&=&-\int \dfrac{d^3kd^3q}{(2\pi)^6}(2\pi kq)\delta \phi_\bfk \delta \phi_\bfq\int_{0}^{r_{W}(t)} rdr \Big[\dfrac{2\cos\psi\cos(\alpha-\psi)}{|\bfk+\bfq|r}\sin(|\bfk+\bfq|r)  \nn\\
&&+\Big(\cos\psi\cos(\alpha-\psi)-\dfrac{1}{2}\sin\psi\sin(\psi-\alpha)\Big)  \nn\\ 
&&\times\Big(\dfrac{-4\sin(|\bfk+\bfq|r)}{|\bfk+\bfq|^3r^3}+\dfrac{4\cos(|\bfk+\bfq|r)}{|\bfk+\bfq|^2r^2}\Big)\Big]  \nn\\ 
&&=-\int \dfrac{d^3kd^3q}{(2\pi)^6}(2\pi kq)\delta \phi_\bfk \delta \phi_\bfq\Big[ \dfrac{-2}{|\bfk+\bfq|^4}(2kq+(k^2+q^2)\cos\alpha)  \nn\\ 
&&-\dfrac{2\cos(|\bfk+\bfq|r_{W}(t))}{|\bfk+\bfq|^4}(kq+kq\cos^2\alpha+(k^2+q^2)\cos\alpha)  \nn\\ 
&&+\dfrac{4\sin(|\bfk+\bfq|r_{W}(t))}{|\bfk+\bfq|^5 {r_W(t)}}\Big(\dfrac{3}{2}kq+\dfrac{1}{2}kq\cos^2\alpha+(k^2+q^2)\cos\alpha\Big)\Big]
\ea

After combining all contributions, the interaction Hamiltonian becomes 
\ba
H_I(\eta)&=&2\epsilon \theta(\eta-\eta_0)a(\eta)^2\int \dfrac{d^3kd^3q}{(2\pi)^6}(\mathcal H_I^{(1)}+\mathcal H_I^{(2)}+\mathcal H_I^{(3)}) \, ,
\ea
where 
\ba
\mathcal H_I^{(1)}&=&\dfrac{2\pi}{\K^3}\partial_{\eta}\delta \phi_\bfk \partial_{\eta}\delta \phi_\bfq \times \nn \\ 
&&\Big[ \dfrac{\beta \K}{2H}(1-\cos(\K r_{W}))+\K r_{W} \cos(\K r_{W})-\sin (\K r_{W})\Big]\, \\ 
\mathcal H_I^{(2)}&=&-\dfrac{\pi kq \beta}{H\K^4}\delta \phi_\bfk \delta \phi_\bfq \Big[ 2kq+(k^2+q^2)\cos \alpha \nn \\ 
&&+\cos (\K r_{W}) \big(kq+kq\cos^2\alpha+(k^2+q^2)\cos\alpha \big) \nn \\ 
&&-\dfrac{2\sin (\K r_{W})}{\K r_{W}} \big(\dfrac{3}{2}kq+\dfrac{1}{2}kq\cos^2\alpha+(k^2+q^2)\cos \alpha\big )\Big] \, ,  \\
\mathcal H_I^{(3)}&=&\dfrac{-2\pi kq\cos\alpha}{\K^3}\delta \phi_\bfk \delta \phi_\bfq \Big[ \sin(\K r_{W})-\K r_{W} \cos(\K r_{W})\Big] \, .
\ea

We remind that in calculating the above integrals, and in the limit that we neglect the gravitational 
backreactions,  we use the profile of a massless scalar field for the inflaton field 
\ba
\delta \phi_{\bfk} = \frac{H}{\sqrt{2 k^{3}}} ( 1+ i k \eta) e^{-i k \eta} \, .
\ea

\subsection{Corrections in inflaton's two point function }

The form of the interaction Hamiltonian given above is too complicated to calculate the integral in Eq. (\ref{Dyson}) analytically.  Instead, here we look at various interesting limits in which the in-in integral can be performed analytically. 

First, let us  check if  there is any strong  amplification in two point function in the limit 
$\bfk+\bfq\to 0$. This configuration arises if the translation invariance holds, 
requiring $\bfk+\bfq= 0$  from the momentum conservation. This configuration contributes to the diagonal components of the CMB temperature  power spectrum.  
Of course, in the presence of the bubble the translation invariance is lost so there is no requirement of having $\bfk+\bfq = 0$.  
Interestingly, all of the apparent singularities in $H_I$ in the limit  $\bfk+\bfq \to 0$ cancels, yielding 
\ba
\lim_{\bfk+\bfq\to 0}\mathcal H_I=k^2\delta \phi_\bfk\delta \phi_\bfq \dfrac{\pi r_{W}^3}{3}(2-\dfrac{r_f}{r_{W}})+2\pi r_{W}^2 \delta \phi_\bfk' \delta \phi_\bfq' (\dfrac{r_f}{2}-\dfrac{r_{W}}{3}) \, ,
\ea
in which, $r_f$ is the final asymptotic comoving radius of the bubble. 

Correspondingly, the result of the in-in integral, including the time dependence of $r_{W}(t)$,  in this limit is obtained to be  
\ba
\label{diagc}
\lim_{\bfk+\bfq\to 0} \Delta \langle \delta \phi_\bfk \delta \phi_\bfq \rangle &=&-\dfrac{\pi \epsilon H^2 r_f}{6k^5}(6r_f^2k^2+25)-\dfrac{\pi \epsilon H^2 r_f}{6k^5}(2r_f^2k^2+3)\, , \nn \\ 
&=&\dfrac{-2\pi \epsilon H^2 r_f^3}{3k^3}(2+\dfrac{7}{k^2r_f^2}) \, .
\ea

Now we focus on the off diagonal elements. To simplify the analysis further, we assume configurations in which the size of CMB sphere is much smaller than the final size of the bubble. In such cases the CMB observer can probe only the small modes with  $|\bfk+\bfq|r_f\gg 1$. In this limit, the Hamiltonian simplifies to 
\ba
\label{Int}
\lim_{|\bfk+\bfq| r_{f}\gg 1}\mathcal H_I&=&\dfrac{2\pi}{|\bfk+\bfq|^2}\delta \phi_\bfk'\delta \phi_\bfq' \Big((r_{W}-r_f)\cos(\K r_{W})+r_f\Big) \nn\\ 
&&-\dfrac{{2}\pi kq r_f}{\K^4}\delta \phi_\bfk \delta \phi_\bfq \Big(2kq+(k^2+q^2)\cos\alpha+ \nn\\ 
&&\cos(\K r_{W})[kq+kq\cos^2\alpha+k^2\cos\alpha+q^2\cos\alpha]\Big) \nn\\ 
&&+\dfrac{{2\pi} kq\cos \alpha }{\K^2} r_{W}(\tau)\cos (\K r_{W})\delta \phi_\bfk \delta \phi_\bfq \, .
\ea
Even with these simplifications, the result of the in-in integral is too complicated to report for a general shape. However, there is a window of momenta in which the power spectrum may peak sharply. Let us define $K\equiv \K$.  Using \eqref{Int} one can easily see that when $K=k+q$ there is a potential place for resonance between the expression $\exp(ik\tau)\exp(iq\tau)$  inside the in-in integral and the 
classical behavior $\cos (K (r_f-\tau))$  in the interaction Hamiltonian. In this limit, we only keep the  oscillatory cosine term in $H_I$ and neglect the other terms. In addition, we also neglect the mild time dependence in coefficient of the oscillatory terms. With these simplifications the oscillatory part of the 
two point function is calculated to be 
\ba
\label{inin}
\Delta \langle \delta \phi_\bfk \delta \phi_\bfq \rangle^{\mathrm{osc}} &=&\dfrac{{-2\pi  H^2\,  \epsilon} \, r_f\sin^2\alpha}{K^4 k q }\dfrac{\cos(Kr_f)}{(K+k+q)(K-k-q)} \times \nn \\ 
&& \Big[K(K+k+q)(K-k-q)ln(\dfrac{K+k+q}{K-k-q})+2(k+q)(k^2+q^2+kq-K^2)\Big]  \, ,\nn\\ 
&&=\dfrac{ {-2\pi \epsilon} H^2 r_f \sin^2\alpha}{kqK^4}\cos (Kr_f) \nn\\ 
&&~~~~~~~~\times \Big[ K\ln (\dfrac{K+k+q}{K-k-q})+\dfrac{2(k+q)(k^2+q^2+kq-K^2)}{K^2-(k+q)^2}\Big]\, .
\ea
From this calculation we see that the aforementioned resonance (in $\alpha\to 0$ limit) disappears. 
The high oscillatory term $\cos(Kr_f)$  in the limit $K r_f\gg 1$ can not be taken too seriously. The reason is that this high oscillatory pattern is the artifact of our assumption that the the bubble wall has no thickness. Considering a realistic situation in which the bubble wall has a finite thickness, then rapid oscillations
for the  wavelength  shorter than the width of the shell thickness  disappear.

The non-oscillatory contribution to two point function is  given by
\ba
\label{non-osc}
\Delta \langle \delta \phi_\bfk \delta \phi_\bfq \rangle^{\mathrm{non-osc}}=\dfrac{-4\pi \epsilon r_f H^2}{K^2(k+q)kq}+\dfrac{4\pi \epsilon r_f H^2}{k^2q^2 K^4(k+q)}(k^2\cos\alpha+q^2\cos\alpha+2kq)(k^2+q^2+kq) .
\ea

Comparing \eqref{inin} and \eqref{non-osc} with \eqref{diagc} we observe that in the limit  $|\bfk r_f|\gg 1$, the diagonal elements of the two point function in Fourier space induced by the the bubble scales like $(kr_f)^3$ while the off-diagonal terms scales like $kr_f$. Naively this suggests that for these (short) modes the approximate homogeneity holds and the off-diagonal terms can be neglected to leading order. 
However, one should note that the leading $r_f^3$ in Eq. \eqref{diagc} can not be distinguished from the overall  isotropic  COBE normalization.  To see this,  suppose that we quantize the mode $\bfk$ in a box with the size $r_f$. In this box we can replace $(2\pi)^3\delta^3(\bfk+\bfq)$ with $r_f^3$. So the total Power spectrum is 
\be
{\cal P}_{\bfk}={\cal P}_0\Big(1-\dfrac{28\pi \epsilon}{3k^2r_f^2}\Big) \, ,
\ee
where ${\cal P}_0=(\dfrac{H^{2}}{2 \pi\dot{\phi}})^2 (1-\dfrac{8\pi \epsilon}{3})$  is the isotropic power spectrum. As just mentioned, we see that the leading $r_f^3$ correction in Eq. \eqref{diagc} only modifies 
the normalization of ${\cal P}_0$. Therefore, the observable scale-dependent corrections  in power spectrum  is given by the subleading term in Eq. \eqref{diagc} which  induces correction  in ${\cal P}_\bfk$
scaling  like $1/(k r_{f})^{2}$.

The above formula suggests that the  corrections in diagonal components of the curvature perturbation two point function is negative and falls off rapidly on small scales. In addition,  the off-diagonal corrections to the two point function are at the same order,  $1/(kr_f)^2$. Therefore,   the effect of the bubble can be viewed as due to violation of homogeneity. A careful CMB data analysis is required to study the predictions of this model and to see if the contributions of the bubble to diagonal and off-diagonal components improve the fit to the data which can also be  used  to constrain the model parameters.

\subsection{Total curvature perturbations}

In the previous section, we have calculated everything in terms of the exterior coordinate 
$(t_+, r_+)$ as the analysis were simpler. However, once the bubble wall sweeps the CMB sphere, the 
coordinate which is naturally available to the observer is the interior coordinate $(t_-, r_-)$. Therefore, we have to be careful in identifying the true curvature perturbations for an observer who enters the bubble. More specifically, a change into the interior coordinate will induce a change in $\delta \phi$. This in turn induces a shift in curvature perturbations as given in  Eq. (\ref{R-eq}). Here we calculate the corrections in curvature perturbations upon this coordinate transformation. 

We remind that the relation between the exterior coordinate and the interior coordinate, as obtained in section \ref{relation} is given by  
\ba
t_-&=&\big(1+\epsilon \theta(r_f-r) \big)t \nn \\ 
r_-&=&\big(1+\epsilon \theta(r_f-r) \big)r-\epsilon \theta(r_f-r) r_f \, .
\ea
Under this transformation the inflaton fluctuation changes to 
\be
{\delta \tilde \phi}(t_e,r_-)=\delta \phi(t_e,r)+\epsilon \theta (r-r_f) (r-r_f)\partial_r \delta \phi
\ee
Note that since we make the measurement at the time of  end of inflation when the super-horizon  modes are frozen, therefore  there is no time derivative of $\delta \phi$ involved. 

Correspondingly, the total curvature perturbations on comoving slices from Eq. (\ref{R-eq}) becomes
\be
\calR=-\dfrac{H}{\dot{\phi}}\delta {\tilde \phi}=-\dfrac{H}{\dot{\phi}}\Big(\delta \phi+\epsilon \theta (r-r_f) (r-r_f)\partial_r \delta \phi\Big) \, .
\ee
Hence, in Fourier space
\be
\calR_\bfk=-\dfrac{H}{\dot{\phi}}\Big( \delta\phi_{\bfk}+ \epsilon \int_{0}^{r_f}(r-r_f) d^3x\int \dfrac{d^3Q}{(2\pi)^3} \delta \phi_\bfQ (i\bfQ.\hat{x})\exp(i(\bfQ-\bfk).\bfx)\Big) \, .
\ee
From the above expression we see that the correction in curvature perturbation from this coordinate transformation is at the order $\epsilon$. Therefore,  the correction in total curvature perturbation power spectrum to leading order in $\epsilon$ becomes a simple integral (not a nested integral like in  previous section).  Consequently, the contribution of this field redefinition in  curvature perturbation power spectrum, denoted by $\Delta_{\mathrm{red}}\langle\calR_\bfq\calR_\bfk\rangle $,  becomes
\ba
\label{power-red}
\Delta_{\mathrm{red}}\langle\calR_\bfq\calR_\bfk\rangle &=&2\pi \epsilon  \big(\dfrac{H^2}{\dot{\phi}} \big)^2 \cos(\alpha-\psi)   \dfrac{Kr_f\cos Kr_f+2Kr_f-3\sin Kr_f}{q^2K^4}+k\leftrightarrow q \\ \nn
&=&2\pi \epsilon \big(\dfrac{H^2}{\dot{\phi}} \big)^2\dfrac{Kr_f\cos Kr_f+2Kr_f-3\sin Kr_f}{K^5 k^2q^2}
 (k+q)(kq+\cos\alpha [k^2+q^2-kq]) \, .
\ea
As an estimation of this contribution, we note for example that  in the limit $|\bfk+\bfq|\to 0$ this term vanishes so it does not contribute to the diagonal components of the CMB power spectrum. However, 
it contributes to the off-diagonal terms for small scales $Kr_f\gg 1$ with the same order of magnitude as  in Eqs. \eqref{inin} and (\ref{non-osc}), scaling like $ k r_f$.

The total correction in curvature perturbation power spectrum induced by  the bubble  is the 
sum of Eqs. (\ref{power}) and (\ref{power-red}). 

\section{Discussions}

In this work we have studied the imprints of a bubble, formed from the decay of the false vacuum into the true vacuum, in curvature perturbation power spectrum. 
This question is interesting from various points of view. The topic of vacuum decay and tunneling in field space has captured significant interests in past decades which also influenced the development of inflationary scenarios. Therefore, it is an interesting question to look for the observational imprints of the bubble formation from the decay of false vacuum. On a more speculative side, this may also open a new observational window to test the landscape scenarios in connection to CMB observations.

On the other hand, there are hints of anomalies in CMB maps, specially on low $\ell$ parts of CMB power spectrum. Among these anomalies are power asymmetry and statistical anisotropies. Although the statistical significances of these anomalies are not high, but looking for theories which can explain these anomalies is an interesting exercise. As we have seen, the formation of bubble breaks the translation invariance while the spacetime is still isotropic. In addition, the bubble is expanding into the false vacuum region sweeping more and more regions into its interior. An observer at the end of inflation who looks at a particular mode which has left the horizon around the time when the bubble has crossed the CMB will observe corrections in CMB power spectrum which are anisotropic with non-trivial scale dependence.

In order to perform the analysis, we have considered a simple case where the bubble has no thickness, i.e. the limit of thin wall approximation studied in  \cite{Coleman:1977py, Coleman:1980aw}. In addition, we have assumed that the difference between the false vacuum and the true vacuum potentials, measured by the parameter $\epsilon$, is small so we can calculate the corrections in curvature perturbations perturbatively to leading order in $\epsilon$. In addition, in order not to produce large quadrupole and octuple anisotropies from the classical inhomogeneous modification of the inflaton field, we require $\beta \ll1$, corresponding to a small initial size of the bubble at the time of formation. In this limit the effects if gravity is small, and one is effectively dealing with a bubble in Minkowski space as in \cite{Coleman:1977py}. The bubble expands quickly and soon reaches to its  asymptotic relativistic speed. 

There are various directions that the current analysis can be further improved. One direction is to look for the evolution of the background inflaton field in the presence of the bubble. For this, one has to solve the 
scalar field equation with the appropriate boundary conditions on the surface of bubble. Having obtained the profile of the background field, one can extend the current analysis and look for the corrections in curvature perturbation power spectrum. The other direction is to compare the predictions of the scenario to a full CMB  data and perform the likelihood analysis to put constraints on model parameters.  For this purpose we need either to compute the correlation functions in real space and (or) compute $C_{ll'mm'}$ for temperature fluctuations. These are interesting directions which are beyond the scope of the current work. 
\\ \vspace{0.5cm}


{\bf Acknowledgments:}  
We thank Jaume Garriga, Hideo Kodama, Misao Sasaki and Takahiro Tanaka  for very useful discussions. H. F. would like to thank Yukawa Institute for Theoretical Physics (YITP) and  the University of Barcelona 
for hospitality where this work was in progress. S. J. would like to thank YITP for hospitality during the development of this work.


\end{document}